\documentstyle[12pt,epsf,rotate]{article}

\makeatletter
\renewcommand{\@cite}[2]{$^{\mbox{\scriptsize #1\if@tempswa,#2\fi)}}$}
\renewcommand{\@biblabel}[1]{#1) }
\makeatother

\def\simle{\mathrel{\rlap{\raise 2pt \hbox{$<$}}
    {\lower 2pt \hbox{$\sim$}}}} 
\def\simge{\mathrel{\rlap{\raise 2pt \hbox{$>$}}
    {\lower 2pt \hbox{$\sim$}}}} 
\def\lambdacut{\mathrel{\rlap{\raise2pt\hbox{/}}
    \lower1pt\hbox{$\lambda$}}}   
\def\lambdabar{\mathord{\rlap{\raise2pt\hbox{\hskip2pt -}}
    \hbox{$\lambda$}}}   
\def\ket#1{| {#1} \rangle}
\def\bra#1{\langle {#1} |}
\def\d{{\rm d}}

\def\ie{{\it i.e.\ }}
\def\eg{{\it e.g.\ }}
\def\viz{{\it viz.\ }}
\def\etal{{\it et al.}}

\long\def\drafton{\bgroup\tt}
\long\def\draftoff{\egroup}
\def\forcenewpage{\vfill\eject}
\def\forcenewpage{}

\begin{document}


\begin{titlepage}
\vspace*{1cm}
\makebox[2cm]{}\\[-1in]
\begin{flushright}
\begin{tabular}{l}
hep-ph/9605290\\
April 1996
\end{tabular}
\end{flushright}
\vskip2cm
\begin{center}
{\Large\bf Phenomenological Introduction to \\[4pt]
Direct Dark Matter Detection \\}

\vspace{2cm}
Paolo Gondolo\\
\vspace{0.7cm}
{\it University of Oxford, Department of Physics, Theoretical Physics,\\ 
  1 Keble Road, Oxford, OX1 3NP, United Kingdom}
\vfill

{\bf Abstract\\[5pt]}
\parbox[t]{\textwidth}{
The dark matter of our galactic halo may be constituted by elementary
particles that interact weakly with ordinary matter (WIMPs). In spite
of the very low counting rates expected for these dark matter
particles to scatter off nuclei in a laboratory detector, such direct
WIMP searches are possible and are experimentally carried out at
present. An introduction to the theoretical ingredients entering the
counting rates predictions, together with a short discussion of the
major theoretical uncertainties, is here presented.
}

\vspace*{5cm}

{\em invited talk at the XXXI Rencontres de Moriond,\\
``Dark Matter in Cosmology, Quantum Measurements, Experimental
Gravitation,'' \\ Les Arcs, France, January 1996
}
\end{center}
\end{titlepage}

\newpage



This is a phenomenological introduction to the detection of dark
matter through its scattering in a laboratory detector. For dark
matter in the form of massive quasi-stellar objects, like brown
dwarfs, which are much bigger and much heavier than the Earth, this
type of detection is quite impracticable if not undesirable. I
therefore consider dark matter in the form of elementary particles.

Many particles, most of which hypothetical, are at present candidates
for dark matter: neutrinos, neutralinos, axions, etc. The methods
employed in hunting for these particles are very different. In this
short note I focus on this meeting's category of particle dark matter,
\viz weakly interacting massive particles or WIMPs.

WIMPs, in a broad sense, are particles with masses of the order of
atomic masses or higher ($m \simge 10 \mbox{GeV}/c^2$) that interact
with ordinary matter with cross sections typical of the weak
interaction or smaller ($\sigma \simle 10^{-38} \mbox{cm}^2$ off a
proton). The presently most popular WIMP is the yet-undetected
neutralino, the lightest supersymmetric particle in supersymmetric
models. Other famous WIMPs are Dirac and Majorana neutrinos, which
however, thanks to the on-going dark matter searches complemented by
accelerator results, we know not to be the dominant component of our
galactic halo.

A general introduction to dark matter has been given by Olive at this
meeting. Direct detection of WIMPs was first explored by Goodman and
Witten.\cite{GoodmanWitten} General reviews are Primack et
al\cite{pss} and Smith and Lewin.\cite{sl} Engel
\etal\cite{EngelReview} present the nuclear physics involved. At this
meeting Cabrera discusses experimental aspects of direct dark matter
detection, while I focus on the theoretical aspects.

It is worth recalling some properties of the dark halo of our
galaxy. Even if recent observations might change the details of our
picture, the 1981 model by Caldwell and
Ostriker\cite{CaldwellOstriker} is good for my purposes. The Sun lies
at a distance of $\approx 8.5$ kpc on the disk of our spiral galaxy,
and moves around the center at a speed of $\approx 220$ km/s. The
luminous disk extends to $\approx 12$ kpc, and is surrounded by a halo
of $\approx 100$ kpc where globular star clusters and rare subdwarf
stars are found. Dynamical arguments suggest that the halo is filled
with dark matter, whose local density in the vicinity of the Sun is
estimated to be $\rho_{\rm DM} = 0.2\hbox{--}0.4 {\rm GeV}/c^2/{\rm cm}^3$ $ =
0.7\hbox{--}1.4\;10^{-24} {\rm g}/{\rm cm.}^3$ Equilibrium
considerations also give the root mean square velocity of halo
consituents to be 200--400 km/s, not much different from the escape
speed from the galaxy (500--700 km/s). Very little is known on the
mean rotation speed of the halo, and we will assume it does not
rotate. All in all, there is an optimistic factor of 2 uncertainty on
the density and velocity of halo dark matter.

Are there WIMPs in our galactic halo? The scientific way to answer
this question is to detect them. Signals could come from WIMP
annihilation (indirect detection) or WIMP scattering (direct
detection). In the former we search for rare annihilation products
like neutrinos, antimatter or gamma-ray lines. This is reviewed by
Bergstr\"om at this meeting. In the latter, the basic philosophy is to
build a target, wait and count.

The WIMP scattering rate per target nucleus is the product of the WIMP
flux $\phi_\chi$ and of the WIMP-nucleus cross section $\sigma_{\chi
i}$.  For an order of magnitude estimate we take the effective
WIMP-nucleon coupling constant to be Fermi's constant $G_F = 2.3016\;
10^{-19} \hbar^2 c^2 \, {\rm cm/GeV}$, which sets the scale of weak
interactions. We distinguish two cases: (i) the WIMP couples to
nucleon spin, $\sigma_{\chi i} \approx G_F^2 \mu_i^2/\hbar^4 \simle
10^{-34} \mbox{cm}^2 $ ; and (ii) the WIMP couples to nucleon number,
$ \sigma_{\chi i} \approx G_F^2 \mu_i^2 A_i^2/\hbar^4 \simle 10^{-30}
\mbox{cm}^2 $. Here $\mu_i = m_\chi m_i/(m_\chi+m_i)$ is the reduced
WIMP-nucleus mass, and $A_i$ is the atomic number of the target
($\approx 80$ in the numerical examples) . The WIMP flux is
$ \phi_\chi = v {\rho_\chi / m_\chi } \approx { 10^7 \mbox{cm}^{-2}
\mbox{s}^{-1} / ( m_\chi c^2 / \mbox{GeV} ) }, $ for a WIMP density
$\rho_\chi \approx 10^{-24}$ g/cm$^3$ and a typical WIMP velocity $v
\approx 300$ km/s. The resulting scattering rates, taking $m_\chi
\approx 100$ GeV/$c^2$, are of the order of $\simle 1$/kg-day for
spin-coupled WIMPs and of $\simle 10^4$/kg-day for WIMPs coupled to
nucleon number.  These rates are quite small compared with normal
radioactivity background. Therefore the common denominator of direct
experimental searches of WIMPs is a fight against background.

For this we get help from characteristic signatures that we do not
expect for the background. For example, while the Earth revolves
around the Sun, the mean speed of the WIMP ``wind'' varies
periodically with an amplitude of 60 km/s. This leads to a $\approx
10$\% seasonal modulation in the detection rate, with a maximum in
June and a minimum in December.\cite{Drukier} As another example, the
direction of the WIMP ``wind'' does not coincides with the Earth
rotation axis, so the detection rate might present a diurnal
modulation due to the diffusion of WIMPs while they cross the
Earth\cite{Avignone} (this however occurs for quite high cross
sections). A final example of background discrimination is that the
WIMP signal is directional, simply because most WIMPs come from the
direction of the solar motion.\cite{Spergel}

\subsection*{\bf WIMP-nucleus scattering}

Since the relative speed $v\approx 300 \mbox{km/s} \approx 10^{-3} c$,
the process can be treated non-relativistically. The center of mass
momentum is given in terms of the reduced WIMP-nucleus mass as $ k =
\mu_i v $ and is $ \simle A_i \, {\rm MeV}/c $ since $\mu_i \le m_i $. The
corresponding de Broglie wavelength is $\simge 200\, {\rm fm}/A_i$, and
can be smaller than the size of heavy target nuclei, in which case
nuclear form factors are important. In the laboratory frame, the
nucleus recoils with momentum $q = 2 k \sin(\theta_{\rm cm}/2) $ and
energy $\nu = q^2/2m_i$. Here $\theta_{\rm cm}$ is the
center-of-mass scattering angle. The 4-momentum transfer is very
small, $Q^2 \simle A_i^2 \, 10^{-6} \,{\rm GeV}^2/c^2 $ (compare with a
typical deep inelastic $Q^2 \simge 1\,{\rm GeV}^2/c^2 $).  

The differential scattering rate per unit recoil energy and unit
target mass is formally 
\begin{equation}
{\d R \over \d \nu} = {\rho_\chi \over m_\chi} \sum_i f_i \eta_i(q) 
{\overline{\left\vert T_i(q^2) \right\vert^2} \over 2 \pi \hbar^4 }.
\end{equation}
The sum is over the nuclear isotopes in the target, $T_i(q^2)$ is the
scattering matrix element at momentum transfer squared $q^2 = 2 m_i
\nu$, and $f_i$ is the mass fraction of isotope $i$. A sum over final
and average over initial polarizations is understood in
$\overline{|T_i(q^2)|^2}$. The factor
\begin{equation}
\eta_i(q) = \int_{q/2\mu_i}^\infty {f_\chi(v) \over v} \d^3 v,
\end{equation}
with units of inverse velocity, incorporates the $\chi$ velocity
distribution $f_\chi(v)$. For a Maxwellian distribution with velocity
dispersion $v_{\rm rms}$, seen by an observer moving
at speed $v_O$,
\begin{equation}
\eta_i(q) = {1\over 2v_O} \left[ {\rm erf}\left({v_q+v_O \over \sqrt{2}
v_{\rm rms}}\right)-{\rm erf}\left({v_q-v_O \over \sqrt{2}
v_{\rm rms}}\right) \right],
\end{equation}
with $v_q=q/2\mu_i$. For standard halo parameters, $\eta_i(q)$ is
approximately exponential in the deposited energy $\nu$. The
previously-mentioned modulations enter the rate through $\eta_i(q)$.

The scattering matrix element $T(q^2)$ can be written as the Fourier
transform
\begin{equation}
T(q^2) = \int \bra{\rm f}  V({\vec r}) \ket{\rm i} {\rm e}^{i {\vec
q} {\vec r}/\hbar} \d{\vec r} 
\end{equation}
 of a non-relativistic WIMP-nucleus potential
\begin{equation}
V({\vec r}) = \sum_{ {\rm pointlike} \atop {{\rm nucleons} \atop
 n={\rm p,n}} }
\left( G^n_s + G^n_a {\vec \sigma}_\chi {\vec \sigma}_n \right)
\delta({\vec r}-{\vec r}_n) .
\end{equation}
The constants $G^n_s$ and $G^n_a$ are effective four-fermion coupling
constants for nucleon-WIMP interactions, and are analogous to Fermi's
constant $G_F$.  $G_s^n$ represents scalar\footnote{Associated to
scalar and axial vectors under 3d rotations.} or spin-independent
interactions, $G_a^n$ axial$^1$ or spin-dependent interactions. Both
terms are coherent in the quantum-mechanical sense when $q R_{\rm
nucleus} \ll \hbar$, \ie when the nucleus can be treated as pointlike
and $T(q^2)$ can be taken as $T(0)$. At larger $q$, which can occur
with heavy target nuclei, both terms are incoherent. Nuclear form
factors $F(q^2)$, conventionally defined by $T(q^2) = T(0) F(q^2)$,
should then be introduced. The scalar and spin form factors are in
general different, reflecting the difference in the mass and spin
distributions inside the nucleus.

The task of a theoretician is to provide a theoretical estimate of
$T(q^2)$ starting from a particle-physics model.  We accomplish this
by stages, successively finding the WIMP-quark, the WIMP-nucleon and
the WIMP-nucleus effective lagrangians.  Step 1, finding the effective
WIMP-quark lagrangian at small $q^2$, is analogous to going from the
Standard Model to four-fermion interactions. Step 2 requires knowledge of
the quark content of the nucleon, \ie the contributions of different
quarks to the nucleon mass and spin. Step 3 needs a nuclear model to
describe how protons and neutrons are distributed in a nucleus.

This procedure is now illustrated for a Dirac neutrino and for a
Majorana particle, an example of which is the neutralino. 

\subsection*{\bf Dirac neutrino}

Step 1: a Dirac neutrino $\nu$ interacts with a quark q through the
diagram in Fig.~1a. At $q^2 \ll m^2_{\rm Z}$, the Z propagator reduces to
$ig^{\mu\nu}/m^2_{\rm Z}$, and the four-fermion amplitude reads
\begin{equation}
\sqrt{2} G_F \; \bar{\nu} ( v_{\nu} - a_{\nu} \gamma_5 ) \gamma_\mu 
\nu \; \bar{\rm q} ( v_{\rm q} - a_{\rm q} \gamma_5 ) \gamma^\mu
{\rm q} ,
\end{equation}
with $v_{\nu} = a_{\nu} = \frac{1}{2}$, $a_{\rm q} = T_{3\rm q}$ and
$v_{\rm q} = T_{3\rm q} - 2 e_{\rm q} \sin^2 \theta_W$.  Here $\sin^2
\theta_W \simeq 0.23 $ and $e_q$ and $T_{3q}$ are the electric charge
and the third component of the weak isospin of quark q.
For a non-relativistic
neutrino, only the time component of the vector current and the space
components of the axial current survive. The first is spin-independent
($ \bar{\nu} \gamma_0 \nu \propto \nu^\dagger\nu$) and the second
spin-dependent ($ \bar{\nu} \vec{\gamma} \gamma_5 \nu \propto
\nu^\dagger \vec{\sigma} \nu$).

Step 2 for the vector part 
\begin{equation}
\sqrt{2} G_F \, v_{\nu} v_{\rm q} \; \bar{\nu} \gamma_\mu \nu \; \bar{\rm
q} \gamma^\mu {\rm q} ,
\end{equation}
because of vector current conservation, simply amounts to
summing $T_{3q}$ and $e_q$ of the constituent quarks.  For protons and
neutrons one obtains respectively
\begin{eqnarray}
G_s^{\rm p} & = & {G_F\over\sqrt{2}} (1-4\sin^2 \theta_W) v_{\nu} \\
G_s^{\rm n} & = & - {G_F\over\sqrt{2}} v_{\nu} .
\end{eqnarray}
The interaction is mainly with the neutrons since $1-4\sin^2
\theta_W\approx 0$. 

Step 2 for the axial part
\begin{equation}
\sqrt{2} G_F \, a_{\nu} a_{\rm q} \; \bar{\nu} \gamma_\mu \gamma_5 \nu \;
\bar{\rm q} \gamma^\mu \gamma_5 {\rm q} , 
\end{equation}
leads to the four-fermion coupling constants
\begin{eqnarray}
G_a^{\rm p} = \sqrt{2} G_F a_{\nu} \left( a_{\rm u} \Delta {\rm u} + a_{\rm d}
\Delta {\rm d} + a_{\rm s} \Delta {\rm s} \right), \\
G_a^{\rm n} = \sqrt{2} G_F a_{\nu} \left( a_{\rm u} \Delta {\rm d} + a_{\rm d}
\Delta {\rm u} + a_{\rm s} \Delta {\rm s} \right).
\end{eqnarray}
Here $\Delta{\rm q}$ is the fraction of the proton spin carried by
quark q, $\frac{1}{2} \bra{\rm p} \bar{\rm q} \gamma_\mu \gamma_5 {\rm
q} \ket{\rm p} = \Delta{\rm q} s_\mu$. It can be
obtained\cite{EllisKarliner} from data on neutron and hyperon
$\beta$-decay, which give $\Delta{\rm u} - \Delta{\rm d} =
1.2573\pm0.0028$ and $\Delta{\rm u} + \Delta{\rm d} -2\Delta{\rm s} =
0.59\pm0.03$, respectively. The contribution of the strange quark is
$\Delta{\rm s}=0$ in the naive quark model, $\Delta{\rm s} =
-0.11\pm0.03\pm\cdots$ from deep inelastic data, and $\Delta{\rm
s}=-0.15\pm 0.09$ from elastic $\nu{\rm p}\to\nu{\rm p}$ data.

Step 3 for the spin-independent part introduces the nuclear mass form factor
$F_{\rm mass}(q^2)$, and results in
\begin{equation}
\overline{\Bigl| T_s(q^2) \Bigr|^2} = \Bigl| Z G_s^{\rm p} + N G_s^{\rm
n} \Bigr|^2 \, \Bigl| F_{\rm mass}(q^2) \Bigr|^2,
\label{Ts}
\end{equation}
where $N$ ($Z$) is the number of neutrons (protons) in the nucleus.
Neutron scattering off nuclei suggests that $F_{\rm mass}(q^2) \simeq
F_{\rm e.m.}(q^2)$, the electromagnetic form factor. The electric
charge distribution is well-described by a Fermi or Woods-Saxon
form,\cite{Hofstadter} whose Fourier transform is indistinguishable
from the convenient analytic expression\cite{Helm}
\begin{equation}
F_{\rm mass}(q^2) \simeq {3 j_1(qR) \over qR} {\rm
e}^{-\frac{1}{2}(qs)^2} .
\label{Fem}
\end{equation}
The electromagnetic radius $R$ and the surface thickness $s$ can be
obtained by fitting electron scattering data,\cite{Hofstadter} or can
be roughly approximated by $R \approx A^{1/3}$ fm and $s \approx $ 1
fm.\cite{EngelReview} $ F_{\rm e.m.}(q^2)$ presents diffraction zeros when
the modified Bessel function $j_1(qR)=0$, the first of which occurs at
$qR \simeq 4.2$. In electron scattering, these diffraction zeros are
filled in, because due to the long-range Coulomb attraction the
electron wave function is distorted from a simple plane wave and the
form factor is not simply the Fourier transform of the charge
density. The short-range nature of WIMP-nucleus interactions make us
expect no wave function distortion, and diffraction zeros
remain.\footnote{The first corrections are at a level of $10^{-6}$ and
come from neglected higher powers of the incoming WIMP velocity.}  The
first diffraction zero is important in assessing bounds from some
present-day detectors.\cite{Bottino}

\begin{figure}[t!]
\centering
\epsfxsize=\hsize
\epsfbox{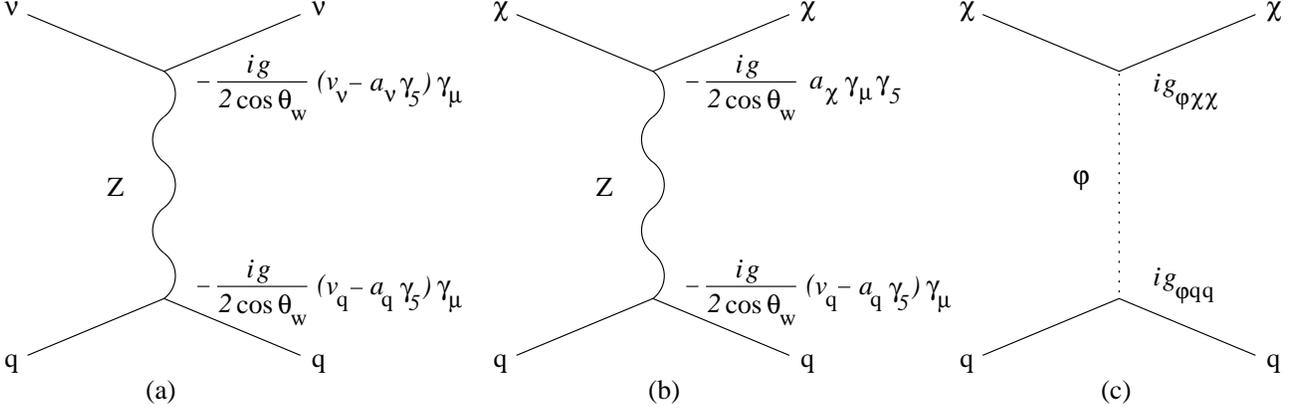}
\caption[]{\small \it Examples of WIMP--quark scattering.}
\end{figure}

Step 3 for the spin-dependent part requires the expectation values of
the total spin of protons $ \langle S_{\rm p} \rangle $ and neutrons
$ \langle S_{\rm n} \rangle $ separately. At $q=0$,
\begin{equation}
\overline{\Bigl\vert T_a(0) \Bigr\vert^2} = {4 (J+1)\over J} \Bigl\vert G_a^{\rm
p} \langle S_{\rm p} \rangle + G_a^{\rm n} \langle S_{\rm n} \rangle
\Bigr\vert ^2 ,
\end{equation}
where $J$ is the nuclear spin.  Even-even nuclei, with even numbers of
protons and of neutrons, do not have spin, and for them $T_a(0)=0$. For
even-odd nuclei with $J\ne 0$, a nuclear model is needed to estimate $
\langle S_{\rm p} \rangle $ and $ \langle S_{\rm n} \rangle $. For
instance, ${}^{73}{\rm Ge}$ is an odd-neutron nucleus with
$J=\frac{9}{2}$. The single-particle shell
model\cite{GoodmanWitten,EllisFlores} gives
\begin{equation}
\langle S_{\rm n} \rangle = {1\over2} \left[ 1 + { \frac{3}{4} -l(l+1)
\over j(j+1) } \right]  = 0.50, \qquad \langle S_{\rm p} \rangle = 0 ;
\end{equation}
the odd-group model,\cite{EngelVogel} in which the odd-nucleon spin is
related to the nuclear magnetic moment $\mu$ and gyromagnetic factors
$g^{L,S}_{\rm n,p}$, gives
\begin{equation}
\langle S_{\rm n} \rangle = { \mu - g^L_{\rm n} J \over g^S_{\rm n} -
g^L_{\rm n} } = 0.23, \qquad \langle S_{\rm p} \rangle = 0 ;
\end{equation}
a more sophisticated interacting shell model\cite{TedRessell} gives
\begin{equation}
\langle S_{\rm n} \rangle = 0.468, \qquad \langle S_{\rm p} \rangle
=0.011.
\end{equation}
The proton might have a small but non-zero contribution to the cross
section, which might change the relative merits of different nuclei
for dark matter searches. 

At $q\ne 0$, nuclear spin form factors are needed. The neutron and
proton contributions differ, and at present only complex
calculations\cite{TedRessell,Fspin} for specific nuclei provide an
estimate of the isoscalar and isovector spin form factors $F_{\rm
spin}^0(q^2)$ and $F_{\rm spin}^1(q^2)$, in terms of which
\begin{equation}
\overline{\Bigl|T_a(q^2)\Bigr|^2} = {J+1\over J} \left| ( G_a^{\rm
p}\!+\!G_a^{\rm n}) \, \langle S_{\rm p}\!+\!S_{\rm n} \rangle \,
F_{\rm spin}^0(q^2) \, + \, ( G_a^{\rm p}\!-\!G_a^{\rm n}) \, \langle S_{\rm
p}\!-\!S_{\rm n} \rangle \, F_{\rm spin}^1(q^2) \right|^2,
\end{equation}
The results of these calculations can be conveniently resumed by the
approximate expressions 
\begin{equation}
F_{\rm spin}^0(q^2) \simeq
\exp\left(-\frac{r_0^2q^2}{\hbar^2}\right), \qquad\qquad
F_{\rm spin}^1(q^2) \simeq \exp\left(-\frac{r_1^2q^2}{\hbar^2}+i\frac{cq}{\hbar}\right) ,\
\end{equation}
with parameters given in the following table for selected nuclei:


\bigskip
\centerline{\vbox{
{ \tabskip=1.1pc \halign{#\hfil &\hfil#\hfil & \hfil#\hfil & \hfil#\hfil
& \hfil#\hfil & \hfil#\hfil & \hfil#\hfil & \hfil#\hfil & \hfil#\hfil \cr
\noalign{\hrule\kern2pt}
 & $J$ & $\nu_{\rm max}$/keV & $\langle S_{\rm p} \rangle$ & $\langle
S_{\rm n} 
\rangle$ & $r_0$/fm & $r_1$/fm & $c$/fm & 
${\rm valid~for} \atop \nu{\rm /keV}<$ \cr
\noalign{\kern2pt\hrule\kern2pt}
${}^{73}$Ge & ${9\over2}$ & 540 & \phantom{-}0.011\phantom{0} & 0.468 & 1.971 & 2.146 &
-0.246 & 55 \cr
\noalign{\kern -3pt}
${}^{28}$Si & ${1\over2}$ & 216 & -0.0019 & 0.133 & 1.302 & 1.548 &
-0.320 & 145 \cr
\noalign{\kern -3pt}
${}^{27}$A & ${5\over2}$ & 100 & \phantom{-}0.3430 & 0.269 & 1.378 & 1.600 &
\phantom{-}0.196 & $\nu_{\rm max}$ \cr
\noalign{\kern -3pt}
${}^{39}$K & ${3\over2}$ & 145 & -0.184\phantom{0} & 0.054 & 1.746 & 1.847 &
\phantom{-}0.371 & $\nu_{\rm max}$ \cr
\noalign{\kern2pt\hrule}
}}}}
\bigskip

\subsection*{\bf Majorana fermion}

A Majorana fermion is a spin-$\frac{1}{2}$ particle that coincides
with its antiparticle. It carries no conserved quantum number. It has
neither vector nor tensor currents. Of the remaining pseudoscalar,
scalar and axial currents, only the last two have a non-vanishing
non-relativistic limit, spin-independent the first ($\bar{\chi} \chi
\propto \chi^\dagger\chi$) and spin-dependent the second ($\bar{\chi} {\vec
\gamma} \gamma_5 \chi \propto \chi^\dagger {\vec \sigma} \chi$).

Axial currents may arise from exchange of a Z boson as in fig.~1b, and
the analysis is then analogous to that in the previous section, with
the obvious replacement of $a_{\nu}$ with $a_\chi$.

Scalar currents originate from exchange of a scalar particle $\varphi$, \eg as in
Fig.~1c. At small $q^2$, the $\varphi$ propagator reduces to $-i/m^2_\varphi$ and
the four-fermion amplitude reads
\begin{equation}
- { g_{\varphi\chi\chi} g_{\varphi \rm qq} \over m^2_\varphi } \; \bar{\chi} \chi \; \bar{\rm
q} {\rm q} .
\end{equation}
For a nucleon $n={\rm p,n}$ one then obtains
\begin{equation}
G_s^n = - {g_{\varphi\chi\chi} \over m^2_\varphi} \, \sum_{\rm q}
g_{\varphi \rm qq} \bra{n}
\bar{\rm q} {\rm q} \ket{n} .
\label{Gs}
\end{equation}
For example, in the case of the neutralino with exchange of the lightest
supersymmetric Higgs boson, the sum over quarks is explicitly
\begin{equation}
{g \over 2 m_{\rm W}} \, \left[ {\cos\alpha\over\sin\beta} \langle m_{\rm
u}\bar{\rm u}{\rm u} + m_{\rm c}\bar{\rm c}{\rm c} + m_{\rm t}\bar{\rm
t}{\rm t} \rangle - {\sin\alpha\over\cos\beta} \langle m_{\rm
d}\bar{\rm d}{\rm d} + m_{\rm s}\bar{\rm s}{\rm s} + m_{\rm b}\bar{\rm
b}{\rm b} \rangle \right] .
\end{equation}
The scalar quark content of the nucleon $\bra{n} \bar{\rm q} {\rm
q} \ket{n}$ can be extracted from data with
the help of chiral perturbation theory, $\pi$-nucleon scattering and
heavy quark expansion.\cite{Cheng} The result is
\begin{eqnarray}
\langle m_{\rm u} \bar{\rm u} {\rm u} \rangle \simeq
\langle m_{\rm d} \bar{\rm d} {\rm d} \rangle \simeq 30 \,{\rm MeV}/c^2, 
\qquad\qquad\quad\!
\langle m_{\rm s} \bar{\rm s} {\rm s} \rangle \simeq 60\hbox{--}120 \,{\rm MeV}/c^2, \\
\langle m_{\rm c} \bar{\rm c} {\rm c} \rangle =
\langle m_{\rm b} \bar{\rm b} {\rm b} \rangle =
\langle m_{\rm t} \bar{\rm t} {\rm t} \rangle =
\frac{2}{27} \left( m_{\rm p} - \sum_{\rm q=u,d,s} 
\langle m_{\rm q} \bar{\rm q} {\rm q} \rangle \right) \simeq
60 \,{\rm MeV}/c^2 .
\end{eqnarray}
The strange quark contribution is uncertain by a factor of 2. Step 3 is
analogous to the Dirac neutrino case, and leads to eq.~(\ref{Ts}) with
four-fermion couplings given by (\ref{Gs}).

\subsection*{\bf Neutralino}

\begin{figure}[t]
\centering
\leavevmode
\epsfxsize=240pt
\epsfysize=240pt
\epsfbox[51 260 506 631]{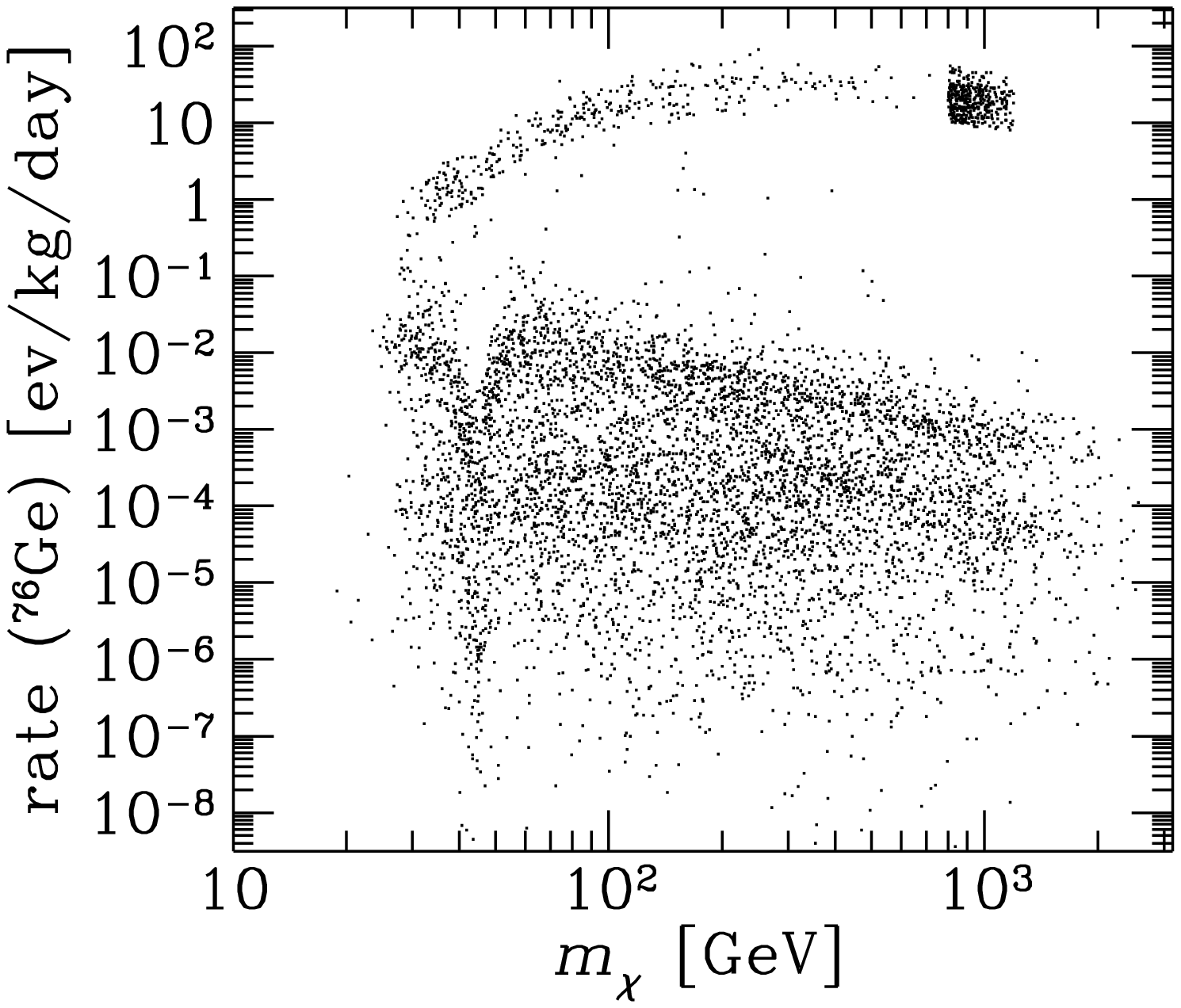}
\leavevmode
\epsfxsize=240pt
\epsfysize=230pt
\rotate[l]{\epsfbox[140 400 470 690]{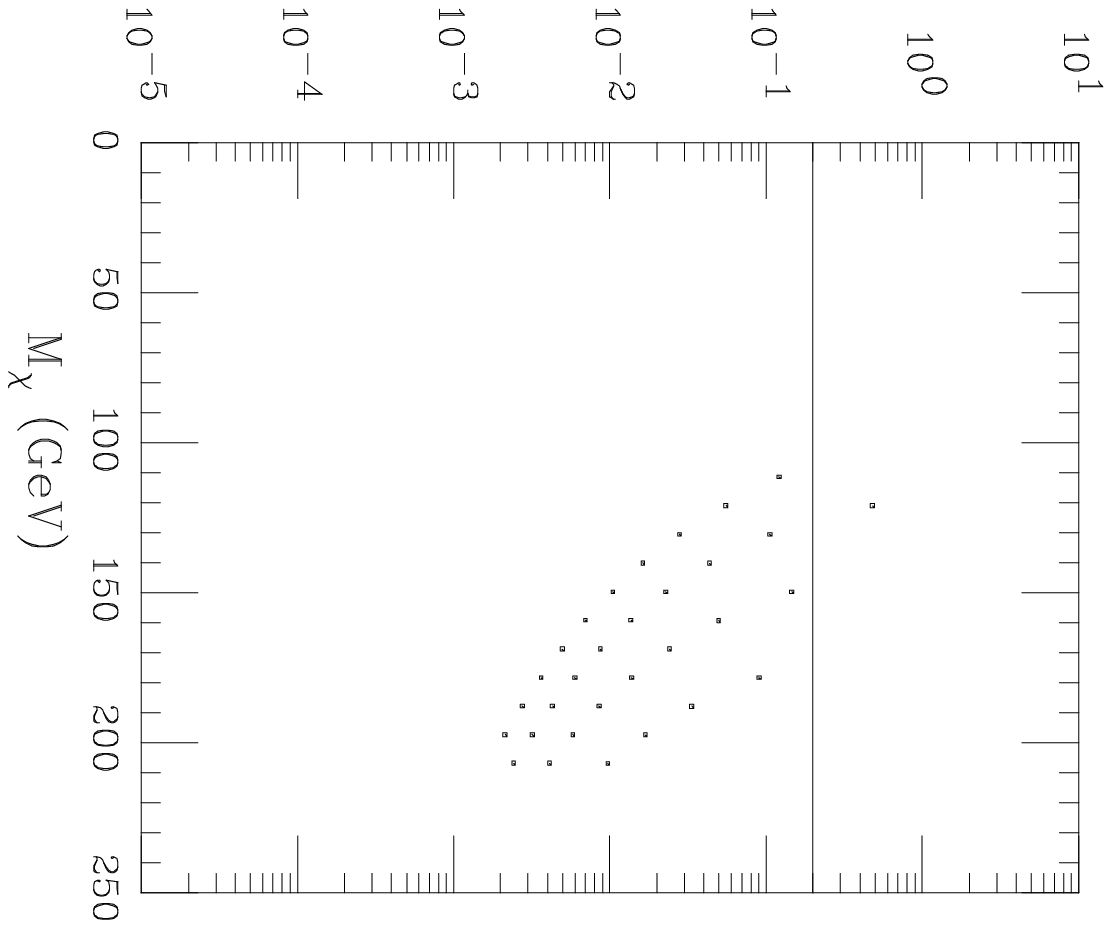}}
\caption[]{\small\it Scattering rate versus mass for neutralinos: (a)
phenomenological approach,\cite{Bergstrom} (b) grand-unified
approach.\cite{Berezinsky}}
\end{figure}

Supersymmetry and the neutralino have been presented by Jungman at
this conference. The neutralino has both spin-dependent and
spin-independent interactions with nuclei, the former mediated by Z
boson and squarks, the latter by Higgs bosons and squarks. The general
formalism of the preceding sections can be used. In the limit of heavy
squarks $\tilde{\rm q}_k$, the effective four-fermion constants are given
by
\begin{equation}
  G_s^{\rm p} \simeq G_s^{\rm n} = \sum_{{\rm q=u,d,s,c,b,t}} \langle
  \bar{\rm q} {\rm q} \rangle \left( - \sum_{h={\rm H}_1,{\rm H}_2} {
  g_{h\chi\chi} g_{h\rm qq} \over m_h^2 } + {1\over 2} \sum_{k=1}^6 {
  g_{L\tilde{\rm q}_k\chi {\rm q}} g_{R\tilde{\rm q}_k\chi {\rm q}}
  \over m^2_{\tilde{\rm q}_k} } \right),
\end{equation}
\begin{equation}
  G_a^{\rm p} = \sum_{{\rm q=u,d,s}} \Delta {\rm q} \left( { g_{{\rm
   Z}\chi\chi} g_{\rm Zqq} \over m_{\rm Z}^2 } + {1\over 8}
   \sum_{k=1}^6 { g_{L\tilde{\rm q}_k\chi {\rm q}}^2 + g_{R\tilde{\rm
   q}_k\chi {\rm q}}^2 \over m^2_{\tilde{\rm q}_k} } \right) , \qquad
   G_a^{\rm n} = G_a^{\rm p}(\Delta{\rm u} \leftrightarrow\Delta{\rm
   d}) .
\end{equation} 
Expressions for the elementary vertices $g_{ijk}$ can be found in
ref.~18.

Predictions in supersymmetric models suffer from the presence of many
unknown parameters. Two extreme attitudes are a
phenomenological approach in which what is not excluded is allowed,
and a grand-unified approach in which coupling constants and masses
are unified at some high energy scale. Fig.~2 shows examples of
calculated event rates in ${}^{76}{\rm Ge}$, each point representing a
choice of model parameters: ``predictions'' may well span 10 orders of
magnitude in a phenomenological approach\cite{Bergstrom} and 2 orders
of magnitude in a more restricted scenario.\cite{Berezinsky}

\forcenewpage
\subsection*{\bf Underabundant dark matter relics}

Given a particle-physics model, the relic density of a species, a WIMP
$\chi$ in particular, is a calculable and definite quantity. Often it
happens that the computed relic density $\Omega_\chi$ is (much)
smaller than the dark matter density in the Universe. For this reason,
some authors simply neglect this case. But even if these WIMPs
constitute only a fraction of the dark matter, they generally have
quite high scattering cross sections off nuclei, because of an
approximate inverse proportionality of the $\chi$ relic density and
the $\chi$--nucleus cross section. However, the scattering rate also
includes the $\chi$ {\it halo} density $\rho_\chi$. It is reasonable
that $\rho_\chi$ is only a fraction of the local dark matter density
$\rho_{\rm DM}$, but which precise fraction it is depends on the model
for galaxy formation. If both the main and the $\chi$ components of
dark matter are cold, we expect them to behave similarly under
gravitation, so that the halo fraction $f_\chi$ might be equal to the
universal fraction $\Omega_\chi/\Omega_{\rm DM}$. Unfortunately,
$\Omega_{\rm DM}$ is poorly known: it can range from $\approx 0.01$
for dark matter associated with galactic halos to $\approx 1$ for a
smooth universal component. In fig.~3, the suppression of scattering
rates due to rescaling of the neutralino halo density by a universal
fraction with $\Omega_{\rm DM} h^2 = 0.025$ is apparent to the left of
the dashed line. This suppression must be included for consistency
when setting bounds on particle-physics models.

\begin{figure}[ht]
\centering
\leavevmode
\epsfxsize=250pt
\epsfbox[54 260 502 600]{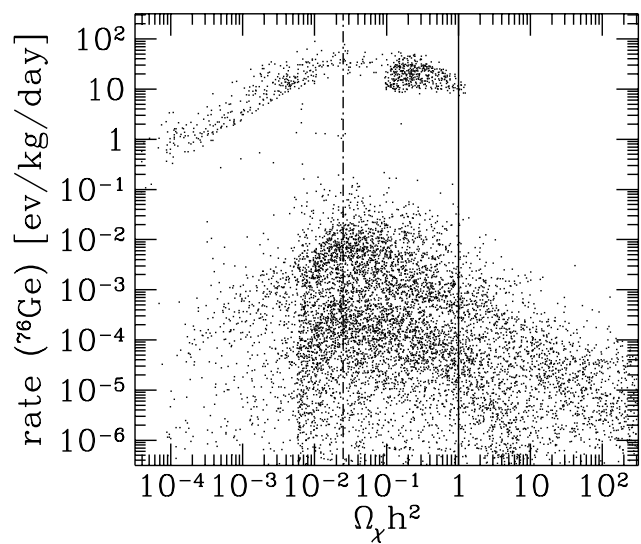}
\caption[]{\small \it Scattering rate versus relic density for
neutralinos (from ref.~18).}
\end{figure}



\end{document}